# Reduction factors for the icosahedral $T_{1u}\otimes h_g$ Jahn-Teller system

Q. C. Qiu,* J. L. Dunn, and C. A. Bates
*School of Physics and Astronomy, University of Nottingham, University Park, Nottingham, NG7 2RD, United Kingdom*

M. Abou-Ghantous
*Department of Physics, CAMS, American University of Beirut, Beirut, Lebanon*

V. Z. Polinger
*Department of Physics, University of Washington, 98195 Seattle, Washington*

Reduction factors (RF's), which are needed when modeling vibronic systems by an effective Hamiltonian in an electronic basis, are calculated for the $T_{1u}\otimes h_g$ Jahn-Teller (JT) system. The results obtained will be useful when modeling the fullerene anion $C_{60}^-$, which is believed to exhibit a $T_{1u}\otimes h_g$ JT effect in its ground state. First-order RF's are calculated using symmetry-adapted vibronic ground and tunneling states in which the system is allowed to tunnel between equivalent minima in the potential energy surface. The effect of anisotropy in the minima is also considered. Second-order RF's are calculated incorporating coupling to excited harmonic-oscillator states associated with the minima.

## I. INTRODUCTION

There has been much work over the last decade concerning the electronic structure and properties of $C_{60}$ molecules and related compounds[1–8] including the vibronic coupling that manifests itself via the Jahn-Teller (JT) effect. The inclusion of vibronic coupling is important because the pattern of energy levels exhibited and the resultant wave functions are different to those that would be expected for a purely electronic system.

Many calculations have been undertaken to determine the energy spectrum of $C_{60}$.[4,5] They show that all levels up to and including the molecular orbital of $H_u$ symmetry are filled and that the lowest unfilled orbital is of $T_{1u}$ symmetry. Therefore the ground state of the anion $C_{60}^-$ must be described by the coupling between the electron in the $T_{1u}$ orbit and the vibrations of the molecular cage. From group theory, couplings to two modes of $a_g$ symmetry and eight modes of $h_g$ symmetry are expected. Although the coupling to the two $a_g$ modes can immediately be distinguished in the spectra, it is a very complex matter to consider all eight $h_g$ modes. A model in which a single $T_{1u}$ orbit interacts with a single $h_g$ mode of vibration in the so-called $T_{1u}\otimes h_g$ JT problem is obviously a good starting point for modeling the real $C_{60}^-$ molecule. This is the main subject of this paper.

In order to interpret the results of spectroscopic experiments such as electron paramagnetic resonance (EPR) or optical absorption/luminescence obtained on any system, it is necessary to determine the pattern of energy levels responsible for producing the lines observed. These can be divided into manifolds that are approximately degenerate but split into quasidegenerate states under the full Hamiltonian. For the purposes of modeling data obtained on JT-active systems, it is common to assume that the basic states form an *electronic* manifold (doublets, triplets, etc.), as these are much simpler to use that the physically correct vibronic states. The vibronic coupling that has been neglected by representing the manifold by orbital states is incorporated into the problem by ''transferring'' the effect of the vibrations into an effective (or spin) Hamiltonian. This includes terms to represent the various perturbations present, such as internal strains, an external stress or spin-orbit coupling.[9–12] Some terms are modifications of terms that would exist without the JT effect and some additional terms are introduced. For example, in magnetic fields one result is that the Landé $g$ factor can take parallel and perpendicular values significantly different from the free-electron value. The coefficients of the terms modified by the inclusion of vibronic coupling involve first-order JT (or Ham) reduction factors (RF's),[13–16] and the coefficients of the additional terms involve second-order RF's. Sometimes further terms are also introduced to explain the observed results even though they do not have any known physical origin.

Effective Hamiltonians, which may implicitly or explicitly make reference to RF's, have been used to help interpret data on a wide range of vibronic systems. The systems most commonly interpreted in this manner involve various ions (usually transition metals or rare earths) in a complex or crystal.[17–27] Other systems that could be interpreted in the manner described above and are currently analyzed by related methods include dilute magnetic semiconductors,[28,29] manganites[30,31] (which are widely believed to show colossal magnetoresistance due to the JT effect), and even the *E. coli* sulfite reductase enzyme.[32] $g$ factors have also been obtained experimentally[33] and theoretically[34] for the $C_{60}^-$ anion, which is of direct relevance to this paper.

In most approaches to interpreting experimental data, the RF's (or parameters involving them, such as $g$ factors) are treated as free parameters whose values are fixed by fitting to experimental data. This can yield useful information on the vibronic coupling. As an example, consider the EPR and optical zero-phonon Zeeman results obtained on chromium-doped GaP. The results can be all explained by modeling the ground state of a $Cr^{3+}$ center as an orbital triplet in which the zero-field spin-orbit coupling is written as

$$-\frac{3}{2}\gamma\lambda\mathbf{l}\cdot\mathbf{s}+\lambda[c(E_\theta E_\theta^s+E_\epsilon E_\epsilon^s)+b(\mathbf{l}\cdot\mathbf{s})^2], \qquad (1)$$

where $\gamma$ is a first-order RF and $b$ and $c$ involve second-order RF's. The $E_\gamma$ and $E_\gamma^s$ are orbital and spin operators, respectively.[35] Fitting to the experimental data gives a best-fit value for $\gamma$ of 0.0121.[35] As this value is much less than 1, this indicates that spin-orbit coupling is almost completely quenched by the JT coupling.[8] The resultant behavior is therefore dominated by the second-order terms $b$ and $c$.[36] This behavior is very different from that obtained neglecting vibronic coupling, when the appropriate description of spin-orbit coupling would be simply $-\frac{3}{2}\lambda\mathbf{l}\cdot\mathbf{s}$ ($-\frac{3}{2}$ is an isomorphic constant that occurs because $|\mathbf{l}|=1$ is a fictitious angular momentum). This confirms that the effect of including the vibronic coupling in an effective Hamiltonian can be very pronounced. However, the values obtained for $\gamma$, $b$, and $c$ by fitting to the experimental data are obtained independently, whereas they are actually related by the JT effect to one coupling constant and one frequency of vibration only.[37] Furthermore, the effects of other ''perturbations,'' such as magnetic field effects, can be expressed in terms of the same constants.

Although the effective Hamiltonian approach as outlined above is useful as a first attempt at modeling experimental data, it is nevertheless unsatisfactory as a complete description of the underlying physical mechanisms. In a fitting procedure, the RF's are treated as free, independent parameters. However, they are not truly free parameters; in a given system the values of the RF's are fixed by the values of the vibronic coupling strength(s) and vibrational frequencies. Further limits on the number of independent parameters are set by sum rule relations between the RF's.[38] In general, the number of free parameters is smaller than the number of RF's. The number of parameters limits the amount of information that can be obtained about the vibronic coupling by experiment.

It is possible to employ various analytical or numerical methods to calculate values for the RF's as functions of the vibronic coupling. These results can then be applied to any given system. Values for the RF's can be determined if the coupling strength is known. Alternatively (as is more likely), it is possible to compare the values for the RF's obtained by fitting with the theoretical results to deduce a value for the strength of the vibronic coupling. Exact quantitative agreement between theoretical values and fitted results will not be obtained because there will always be small additional perturbations present that are not included in the effective Hamiltonian models. However, it can be expected that at least the signs and orders of magnitude of the RF's are in agreement. Thus theoretical calculations are needed to relate the RF's to the coupling strengths. However, the only paper to calculate expressions for RF's in $I_h$ symmetry at present is that by Cullerne *et al.*,[39] who obtain numerical values for the first-order RF's for the $G\otimes g$, $G\otimes h$, and $H\otimes g$ JT systems in the strong coupling limit.

The concept of RF's is based upon the assumption that the ground states with and without the inclusion of vibronic coupling are of the same symmetries. Although this has recently been shown to be not true for certain situations in the strongly coupled icosahedral $H\otimes h$ systems[40,41] and $E\otimes e$ systems with very strong coupling,[42,43] this is generally true and holds for all couplings in the $T_{1u}\otimes h_g$ JT system considered here.

Formally, RF's are defined in such a manner that the matrix elements of the effective Hamiltonian within the electronic basis are the same as those of the perturbation $V$ within the original ground and/or the first excited vibronic states. First- and second-order RF's correspond to the inclusion of $V$ to first or second order in perturbation theory, respectively.[13] In the example above, it was seen that if the first-order RF's are quenched, the second-order terms dominate the observed behavior.[36] Therefore, it is important to calculate expressions for both first- and second-order RF's as a function of the coupling strength in order to determine the relative importance of different contributions.

Second-order RF's are much harder to calculate than those of first-order as they involve coupling to an infinite set of excited states. Nevertheless, results have been found both analytically[44,45] and numerically[46] for many symmetries. In particular, a general method involving the derivation of second-order RF's using symmetry properties of all states, perturbations, and electronic operators was developed first for orbital triplets in cubic symmetry[37] and then for orbital doublet systems.[47] It was shown that second-order RF's can be obtained from the evaluation of the sums of various overlaps of the associated vibrational states. The aim of this paper is to obtain analytical expressions for both first- and second-order RF's for the icosahedral $T_{1u}\otimes h_g$ JT system using the general methods developed previously for cubic systems. The results obtained cover the whole range of coupling strengths. An assumption is that quadratic coupling terms are sufficiently large that the nuclear motion can be treated as being localized around minima in the adiabatic potential energy surface (APES), and not as rotation around a trough. The basis of the transformation method for $T_{1u}\otimes h_g$ and the original results[48] are also summarized.

## II. THE BASIC MODEL

### A. The Hamiltonian and transformation method

Following the work of Fowler and Ceulemans,[49] we formulate the $T_{1u}\otimes h_g$ JT problem with a twofold axis as the $z$ axis, rather than a five-fold axis as used by some other authors, as this gives the most symmetric results. We will label the two components of the $h_g$ mode equivalent to the $e$ modes in cubic symmetry as $\theta$ and $\epsilon$, and the three components equivalent to $t_2$ modes transforming as $yz$, $zx$, and $xy$ as 4, 5, and 6 respectively. $\theta$ and $\epsilon$ are linear combinations of the hydrogenlike $d_{(3z^2-r^2)}$ and $d_{(x^2-y^2)}$ functions.[48]

As the $H$ representation is not simply reducible in $I_h$ symmetry, there are two independent $H$-type quadratic coupling coefficients. There is no unique way of writing down the two different couplings. We will follow the separation with quadratic coupling constants $V_2$ and $V_3$ used by Dunn and Bates.[48] With this separation, the depth of the $D_{5d}$ wells is found to depend upon the $V_2$-type coupling only and the $D_{3d}$ wells upon $V_3$ only. We will therefore write the Hamiltonian in terms of a linear interaction term $\mathcal{H}_1$ and two quadratic terms $\mathcal{H}_2$ and $\mathcal{H}_3$, so that both types of wells can be considered. The total Hamiltonian can then be written in the form $\mathcal{H}=\mathcal{H}_{vib}+\mathcal{H}_1+\mathcal{H}_2+\mathcal{H}_3$, where[48]

$$\mathcal{H}_{vib}=\frac{1}{2}\sum_{\gamma}\left(\frac{P_{H\gamma}^2}{\mu}+\mu\omega^2 Q_{H\gamma}^2\right),$$

$$\mathcal{H}_1=V_1\sum_{\gamma} Q_{H\gamma}C_{H\gamma},$$

$$\mathcal{H}_2=V_2\Bigg[\left(\sqrt{\frac{1}{2}}Q_{H\theta}Q_{H\epsilon}+\sqrt{\frac{3}{8}}(Q_{H4}^2-Q_{H5}^2)\right)C_{H\theta}+\left(\sqrt{\frac{1}{8}}(Q_{H\theta}^2-Q_{H\epsilon}^2+Q_{H4}^2+Q_{H5}^2-2Q_{H6}^2)\right)C_{H\epsilon}$$
$$+\sqrt{\frac{1}{2}}\{\sqrt{3}Q_{H\theta}+Q_{H\epsilon}\}Q_{H4}C_{H4}+\sqrt{\frac{1}{2}}(-\sqrt{3}Q_{H\theta}+Q_{H\epsilon})Q_{H5}C_{H5}-\sqrt{2}Q_{H\epsilon}Q_{H6}C_{H6}\Bigg],$$

$$\mathcal{H}_3=V_3\Bigg[\left(\sqrt{\frac{3}{8}}(Q_{H\theta}^2-Q_{H\epsilon}^2)-\sqrt{\frac{1}{24}}(Q_{H4}^2+Q_{H5}^2-2Q_{H6}^2)\right)C_{H\theta}+\left(-\sqrt{\frac{3}{2}}Q_{H\theta}Q_{H\epsilon}+\sqrt{\frac{1}{8}}(Q_{H4}^2-Q_{H5}^2)\right)C_{H\epsilon}$$
$$+\left(\left(-\sqrt{\frac{1}{6}}Q_{H\theta}+\sqrt{\frac{1}{2}}Q_{H\epsilon}\right)Q_{H4}-\frac{2}{\sqrt{3}}Q_{H5}Q_{H6}\right)C_{H4}+\left(\left(-\sqrt{\frac{1}{6}}Q_{H\theta}-\sqrt{\frac{1}{2}}Q_{H\epsilon}\right)Q_{H5}-\frac{2}{\sqrt{3}}Q_{H4}Q_{H6}\right)C_{H5}$$
$$+2\left(\sqrt{\frac{1}{6}}Q_{H\theta}Q_{H6}-\sqrt{\frac{1}{3}}Q_{H4}Q_{H5}\right)C_{H6}\Bigg]. \quad (2)$$

$V_1$ is the linear vibronic coupling constant, $\mu$ is the reduced mass corresponding to the vibrational mode $Q_{H\gamma}$, and $\gamma$ is summed over all of the modes $\theta$, $\epsilon$, 4, 5, and 6. The $C_{H\gamma}$ are orbital operators, which can be written in the form

$$C_{H\theta}=\frac{1}{2}\sqrt{\frac{3}{5}}\begin{pmatrix}\phi^{-1} & 0 & 0\\ 0 & -\phi & 0\\ 0 & 0 & 1\end{pmatrix},$$

$$C_{H\epsilon}=\frac{1}{2}\sqrt{\frac{1}{5}}\begin{pmatrix}\phi^{2} & 0 & 0\\ 0 & -\phi^{-2} & 0\\ 0 & 0 & -\sqrt{5}\end{pmatrix},$$

$$C_{H4}=\sqrt{\frac{3}{10}}\begin{pmatrix}0 & 0 & 0\\ 0 & 0 & 1\\ 0 & 1 & 0\end{pmatrix},\quad C_{H5}=\sqrt{\frac{3}{10}}\begin{pmatrix}0 & 0 & 1\\ 0 & 0 & 0\\ 1 & 0 & 0\end{pmatrix},$$

$$C_{H6}=\sqrt{\frac{3}{10}}\begin{pmatrix}0 & 1 & 0\\ 1 & 0 & 0\\ 0 & 0 & 0\end{pmatrix} \quad (3)$$

with respect to orbital basis $x$, $y$, and $z$, and where $\phi=\frac{1}{2}(1+\sqrt{5})$ is the golden mean.

We know that the JT effect should result in localization of the nuclear motion about low-symmetry minima in the APES. Previous analyses of the potential energy surface for the $T_{1u}\otimes h_g$ JT problem using numerical[50,51] and analytical methods[52] have shown that, when only linear coupling and harmonic terms are included, there is a continuous spherical equal-energy surface. However, when small anharmonic or quadratic coupling terms are added,[50] which must be present to some extent in a real system, the minimum-energy surface is warped to give either 10 local minima of $D_{3d}$ symmetry or 6 minima of $D_{5d}$ symmetry. Additional points of $D_{2h}$ symmetry can only become absolute minima if coupling to fourth order is included in the Hamiltonian. As this situation only occurs for a very limited range of possible coupling constants, it will not be considered further here.

Mathematically, it is therefore useful to displace the origin of the phonon coordinates to each minimum point in turn. It is then a much simpler matter to describe the vibronic motion about these points. This can be achieved using a method developed originally for tetrahedral systems in order to model magnetic-ion impurities in III-V semiconductors.[53,54] This involves applying a unitary shift transformation

$$U=\exp\left(\mathrm{i}\sum_{\gamma}\alpha_{H\gamma}P_{H\gamma}\right) \quad (4)$$

to displace the nuclear coordinate $Q_{H\gamma}$ to a position $\tilde{Q}_{H\gamma}=Q_{H\gamma}-\alpha_{H\gamma}\hbar$. The resultant transformed Hamiltonian $\tilde{\mathcal{H}}=U^{-1}\mathcal{H}U$ can be split into a contribution $\tilde{\mathcal{H}}_1$ that does not contain any $P_{H\gamma}$ or $Q_{H\gamma}$, and hence does not contain any coupling to excited phonon states, and a second part $\tilde{\mathcal{H}}_2$ that contains all remaining terms. It follows that $\tilde{\mathcal{H}}_1$ will be a good Hamiltonian for determining the ground states of the system in strong coupling. Further work[54] shows that the Hamiltonian is also good for determining excited states. Values for the shift parameters $\alpha_{H\gamma}$ at the minima are therefore found by minimizing the energy of $\tilde{\mathcal{H}}_1$. The result for the $T\otimes h$ problem is minima of $D_{5d}$ symmetry if $\frac{15}{8}\sqrt{2}>3V_2'>\sqrt{5}V_3'>-\frac{15}{8}\sqrt{2}$ and minima of $D_{3d}$ symmetry if $\frac{15}{8}\sqrt{2}>\sqrt{5}V_3'>3V_2'>-\frac{15}{8}\sqrt{2}$, where $V_i'=V_i/\mu\omega^2(i=2,3)$. The $D_{3d}$ wells are labeled $a$ to $j$, and the $D_{5d}$ wells are labeled $A$ to $F$. Anisotropy in the minima can also be included by applying an additional scale transformation.[52,55,56] However, as the excited well states are to be used as approximations to the symmetrized states in the second-order RF calculations and the results are inevitably much more complicated, the anisotropy will only be considered for the first-order RF's in this paper.

The states of the transformed Hamiltonian (i.e., after the application of the shift transformation) are harmonic-oscillator type states representing motion localized about the bottom of a given minimum. These include ground states in the wells and states with a total of $n$ $h$-type phonon excitations, composed of different numbers of the individual component excitations $\theta$, $\epsilon$, 4, 5, and 6. The ground state associated with the well $k$ will be written in this transformed picture in the form $|\psi^{(k)};0>$, where $\psi^{(k)}$ is the orbital state and the 0 indicates that all the $h_g$ oscillators centered on the well are in their ground state.

Vibronic states associated with the wells, such as those determined by the transformation method, are only good eigenstates of Jahn-Teller systems in infinitely strong coupling when the potential barriers separating the wells are infinitely large and no tunneling can take place. This limit is often referred to as the static Jahn-Teller effect. In real systems, the height of the barriers is not infinite and tunneling between equivalent nuclear distortions becomes possible (the dynamic Jahn-Teller effect). Mathematically, this means taking linear combinations of states localized around different wells. As the system is equally likely to be in any one of the equivalent minima, the icosahedral symmetry of the original problem is restored. Therefore, the infinite coupling states should be symmetry-corrected by taking new linear combinations that transform among themselves with the required icosahedral symmetry. Consequently appropriate combinations can be found analytically using projection operator techniques.[57] The results obtained for $T\otimes h$ (Ref. 48) compare well with those of numerical approaches.[58]

As combinations of states localized around different wells are to be taken, it is necessary to write the well states in a common basis. Therefore, states $|\psi^{(k)\prime};n\rangle \equiv|\psi^{(k)\prime};\theta^p\epsilon^q4^r5^s6^t\rangle$ appropriate to the untransformed picture can be obtained by multiplying the transformed states by the value $U^{(k)}$ of $U$ appropriate to that well by substituting with the particular values of $\alpha_j$ for that well. Here, $\theta^p$, for example, denotes $p$ phonon excitations of the $Q_{H\theta}$ mode in the well $k$. As $U^{(k)}$ contains phonon operators, the ground states (with $n=0$) as well as the excited states are automatically vibronic in nature.

An effect of the tunneling is to lift the degeneracy of the well states in finite coupling and restore a three fold $T_{1u}$ ground state. The remaining levels are tunneling levels composed of combinations of localized well states with no phonon excitations in the wells. The energies of these states are higher than the ground state in finite coupling but tend towards the ground state energy as the coupling tends to infinity. The tunneling levels for the $D_{5d}$ wells form a $T_{2u}$ triplet and for the $D_{3d}$ wells they form a $T_{2u}$ triplet and a $G_u$ quartet.

The $x$ components of the $T_{1u}$ ground state and the $T_{2u}$ tunneling state obtained from the $D_{5d}$ minima can be written as

$$\begin{aligned}|0T_{1ux}^{(5d)}\rangle &= N_{T1u}^{(5d)}[\phi^{-1}(|C';0\rangle+|D';0\rangle)+(|E';0\rangle \\ &\quad -|F';0\rangle)],\\ |0T_{2ux}^{(5d)}\rangle &= N_{T2u}^{(5d)}[(|C';0\rangle+|D';0\rangle)-\phi^{-1}(|E';0\rangle \\ &\quad -|F';0\rangle)],\end{aligned} \tag{5}$$

where $N_{T1u}^{(5d)}$ and $N_{T2u}^{(5d)}$ are normalization constants.[48] For the $D_{3d}$ minima, the $x$ components of the symmetry-adapted states are

$$\begin{aligned}|0T_{1ux}^{(3d)}\rangle &= N_{T1u}^{(3d)}[-\phi^2(|c';0\rangle+|d';0\rangle)+(|f';0\rangle-|e';0\rangle)\\ &\quad +\phi(-|g';0\rangle+|h';0\rangle-|i';0\rangle-|j';0\rangle)],\\ |0T_{2ux}^{(3d)}\rangle &= N_{T2u}^{(3d)}[\phi^{-2}(|c';0\rangle+|d';0\rangle)-(|f';0\rangle-|e';0\rangle)\\ &\quad +\phi^{-1}(-|g';0\rangle+|h';0\rangle-|i';0\rangle-|j';0\rangle)],\\ |0G_{ux}^{(3d)}\rangle &= N_{Gu}^{(3d)}[2(|c';0\rangle+|d';0\rangle+|f';0\rangle-|e';0\rangle)\\ &\quad +(-|g';0\rangle+|h';0\rangle-|i';0\rangle-|j';0\rangle)],\end{aligned} \tag{6}$$

where again the $N$'s are normalization constants. We also note that in Ref. 48, the labels $G_{ux}$ and $G_{uz}$ were inadvertently interchanged.

In cubic symmetry, it was possible to derive a full set of symmetry-adapted excited states using projection operators in a similar way to that used to obtain the ground states. However, these are extremely complicated and difficult to evaluate in $I_h$ symmetry. The essential difference is that a group element of the $T_d$ group will transform both the electronic and phonon states associated with one well directly into those for another well, whereas in $I_h$ symmetry the transformation is to a linear combination of states in different wells (with the same overall number of phonon excitations). Therefore, for the second-order RF calculations, we will take the excited states to be excited harmonic oscillator states localized in the wells, rather than symmetry-adapted linear combinations of them. Although these states are only true eigenstates in infinite coupling, the resultant effect of summing over all excited states removes some of the inaccuracies that might otherwise be expected.[59]

## III. REDUCTION FACTORS

### A. First-order reduction factors

Because the $H$ representation in the icosahedral group is nonsimply reducible, additional complications can occur for icosahedral problems involving $H$ (either electronically or vibrationally) that do not occur in other symmetries. This means that it does not automatically follow that derivations developed for systems that are reducible can be applied here. However, it is found that the calculation of first-order RF's in icosahedral symmetry can be carried out in an analogous manner to that used previously.[13] Thus, in first order, the real Hamiltonian for a perturbation of symmetry $\Gamma$ is written in general terms as

$$\mathcal{H}^{(1)}(\Gamma)=\sum_{\gamma} W_{\Gamma\gamma}C_{\Gamma\gamma}, \tag{7}$$

where the $W_{\Gamma\gamma}$ are coefficients and

$$C_{\Gamma\gamma}=\sum_{\gamma_1\gamma_2}\langle\Gamma\gamma_1\Gamma\gamma_2|\Gamma\gamma\rangle|\Gamma\gamma_2\rangle\langle\Gamma\gamma_1| \quad (8)$$

are general expressions for the orbital operators in terms of Clebsch-Gordon (CG) coefficients.[49] A first-order effective Hamiltonian corresponding to the same perturbation $V$ acting between a vibronic state of symmetry $\Gamma_l$ and a state of symmetry $\Gamma_m$ can be written in the form

$$\mathcal{H}_{eff}^{(1)}(\Gamma)=\sum_{\gamma} W_{\Gamma\gamma}K_{\Gamma_l\Gamma_m}^{(1)}(\Gamma)C_{\Gamma\gamma}^{\Gamma_l\Gamma_m}, \quad (9)$$

where the terms $K_{\Gamma_l\Gamma_m}^{(1)}(\Gamma)$ multiplying the electronic operators are defined to be the general first-order RF for a perturbation of symmetry $\Gamma$. For the $T_{1u}\otimes h_g$ problem, both $\Gamma_l$ and $\Gamma_m$ can be taken as $T_{1u}$, $T_{2u}$, or $G_{u,}$. When the effective Hamiltonian in Eq. (9) describes the effect of the perturbation between basis states of the same symmetry as each other, the orbital operators $C_{\Gamma\gamma}^{\Gamma_l\Gamma_m}$ are the same as the $C_{\Gamma\gamma}$ in Eq. (8). When they occur between different states, they have a similar form to Eq. (8), as given explicitly in Eq. (2) of Ceulemans and Chibotaru.[60]

Expressions for the RF's as a function of the coupling strength can be obtained using vibronic states with the required transformation properties and evaluating matrix elements of any operator of the desired symmetry. Suitable states are those given in Eqs. (5) and (6). The matrix elements of the effective Hamiltonian between two general electronic states $|\Gamma_l x_i\rangle$ and $|\Gamma_m x_j\rangle$ are given by

$$\langle\Gamma_l x_i|\mathcal{H}_{eff}^{(1)}(\Gamma)|\Gamma_m x_j\rangle=\sum_{\gamma} W_{\Gamma\gamma}K_{\Gamma_l\Gamma_m}^{(1)}(\Gamma)\times\langle\Gamma_l x_i|C_{\Gamma\gamma}^{\Gamma_l\Gamma_m}|\Gamma_m x_j\rangle. \quad (10)$$

Also, the matrix element of the perturbation Hamiltonian between the corresponding (ground) vibronic states $|0\Gamma_l x_i\rangle$ and $|0\Gamma_m x_j\rangle$ is given by

$$\langle 0\Gamma_l x_i|\mathcal{H}^{(1)}(\Gamma)|0\Gamma_m x_j\rangle=\sum_{\gamma} W_{\Gamma\gamma}\langle 0\Gamma_l x_i|C_{\Gamma\gamma}|0\Gamma_m x_j\rangle. \quad (11)$$

As these two equations must be equivalent, it follows that

$$K_{\Gamma_l\Gamma_m}^{(1)}(\Gamma)=\frac{\langle 0\Gamma_l x_i|C_{\Gamma\gamma}|0\Gamma_m x_j\rangle}{\langle\Gamma_l x_i|C_{\Gamma\gamma}^{\Gamma_l\Gamma_m}|\Gamma_m x_j\rangle}. \quad (12)$$

These expressions can be evaluated using the symmetry-adapted vibronic and orbital states given above.[49] The calculations are simplified noting that a sufficient and necessary condition for nonzero RF's is that the direct product of the symmetries of both vibronic states and the orbital operator must contain $A_{1g}$. It should be noted that although the orbital part of $\mathcal{H}_{eff}^{(1)}$ is neither Hermitian nor anti-Hermitian for some combinations of $\Gamma_l$ and $\Gamma_m$, this approach is valid because the product of the orbital part and its corresponding first-order RF is Hermitian.

As the $T_{1u}\otimes h_g$ JT system is modeled by an electronic state of symmetry $T_{1u}$, all the required symmetry operators $\Gamma$ may be found from the condition that the product $T_{1u}\otimes\Gamma\otimes T_{1u}$ contains $A_{1g}$. This results in the three possible symmetrical operators $A_{1g}$, $T_{1g}$, and $H_g$. Operators of $A_{1g}$ symmetry are trivial and need not be considered here. The calculation of the required matrix elements for $T_{1g}$ and $H_g$ involves evaluating phonon overlaps $\langle 0|U^{(j)\dagger}U^{(k)}|0\rangle$ between wells $j$ and $k$. These are the same factors that appear in the determination of the normalization factors.[48]

Neglecting the anisotropy in the wells and noting that the labels $\Gamma_l$ and $\Gamma_m$ are interchangeable, we find that there are only four distinct nonzero first-order RF's $K_{\Gamma_l\Gamma_m}^{(1)}(\Gamma)$ for $D_{5d}$ wells, namely

$$K_{T_{1u}T_{1u}}^{(1)}(T_{1g})=2X_0S_I,$$

$$K_{T_{1u}T_{1u}}^{(1)}(H_g)=\frac{2}{5}X_0(1+4S_I),$$

$$K_{T_{2u}T_{2u}}^{(1)}(H_g)=0.4,$$

$$K_{T_{1u}T_{2u}}^{(1)}(H_g)=\frac{1}{5}X_0\sqrt{6(1-S_I^2)}, \quad (13)$$

where $X_0=(1+S_I)^{-1}$ and[48] $S_I=\exp[-2(\beta k_1)^2]$ with $\beta=\sqrt{6}/(5-4\sqrt{2}V_2')$ and $k_1=-V_1/(\sqrt{2\hbar\mu\omega^3})$. For the $D_{3d}$ wells, the nonzero RF's are

$$K_{T_{1u}T_{1u}}^{(1)}(T_{1g})=2X_1^2S_D(1+4S_D),$$

$$K_{G_uG_u}^{(1)}(T_{1g})=-\sqrt{\frac{2}{3}}\frac{S_D}{1+S_D},$$

$$K_{T_{2u}G_u}^{(1)}(T_{1g})=\frac{2}{\sqrt{3}}X_2X_3S_D(1-S_D),$$

$$K_{T_{1u}T_{1u}}^{(1)}(H_g)=\frac{2}{5}X_1^2(3+8S_D+14S_D^2),$$

$$K_{T_{2u}T_{2u}}^{(1)}(H_g)=-\frac{2}{5}X_2^2(3-7S_D+4S_D),$$

$$K_{G_uG_u}^{(1)}(H_g)=\frac{\sqrt{2}}{5}X_3^2X_4,$$

$$K_{T_{1u}T_{2u}}^{(1)}(H_g)=\frac{\sqrt{6}}{5}X_1X_2X_4,$$

$$K_{T_{1u}G_u}^{(1)}(H_g)=\frac{4}{5}X_1X_3X_4,$$

$$K_{T_{2u}G_u}^{(1)}(H_g)=\frac{2}{5}X_2X_3(2-3S_D+S_D^2), \quad (14)$$

where $X_1=(3+5S_D+2S_D^2)^{-1/2}$, $X_2=(3-5S_D+2S_D^2)^{-1/2}$, $X_3=(1-S_D^2)^{-1/2}$, and $X_4=(1+S_D-2S_D^2)$, $S_D=\exp[-2(\gamma k_1)^2]$, and $\gamma=\sqrt{2}/(\sqrt{15}-4\sqrt{\tfrac{2}{3}})V_3'$.

In order to interpret these formulas, it is useful to set the quadratic coupling to zero, even though the $D_{3d}$ and $D_{5d}$ points are not actually wells in this case. The RF's with this simplification have been plotted (solid lines) as a function of $k_1$ in Figs. 1 ($D_{5d}$ case) and 2 ($D_{3d}$ case). The nontrivial limits in both strong and weak coupling, as determined analytically from the above expressions, are also marked on the figures. For the $D_{5d}$ minima, it can be seen that all the first-order RF's lie between 0 and 1, but for the $D_{3d}$ minima, some of the $K^{(1)}_{\Gamma_l\Gamma_m}(\Gamma)$ are negative (down to $-2/5$) for $\Gamma_l = \Gamma_m$. $K^{(1)}_{T_{2u}T_{2u}}(H_g)$ for $D_{3d}$ minima is unusual as it changes sign when the coupling strength increases. This implies that the energies of the $T_{2u}$ vibronic states may change much more than other levels as the coupling strength changes from weak to strong. The appearance of negative RF's also shows that the first-order RF's should not be simply regarded as numbers that reduce the effect of an electronic perturbation.

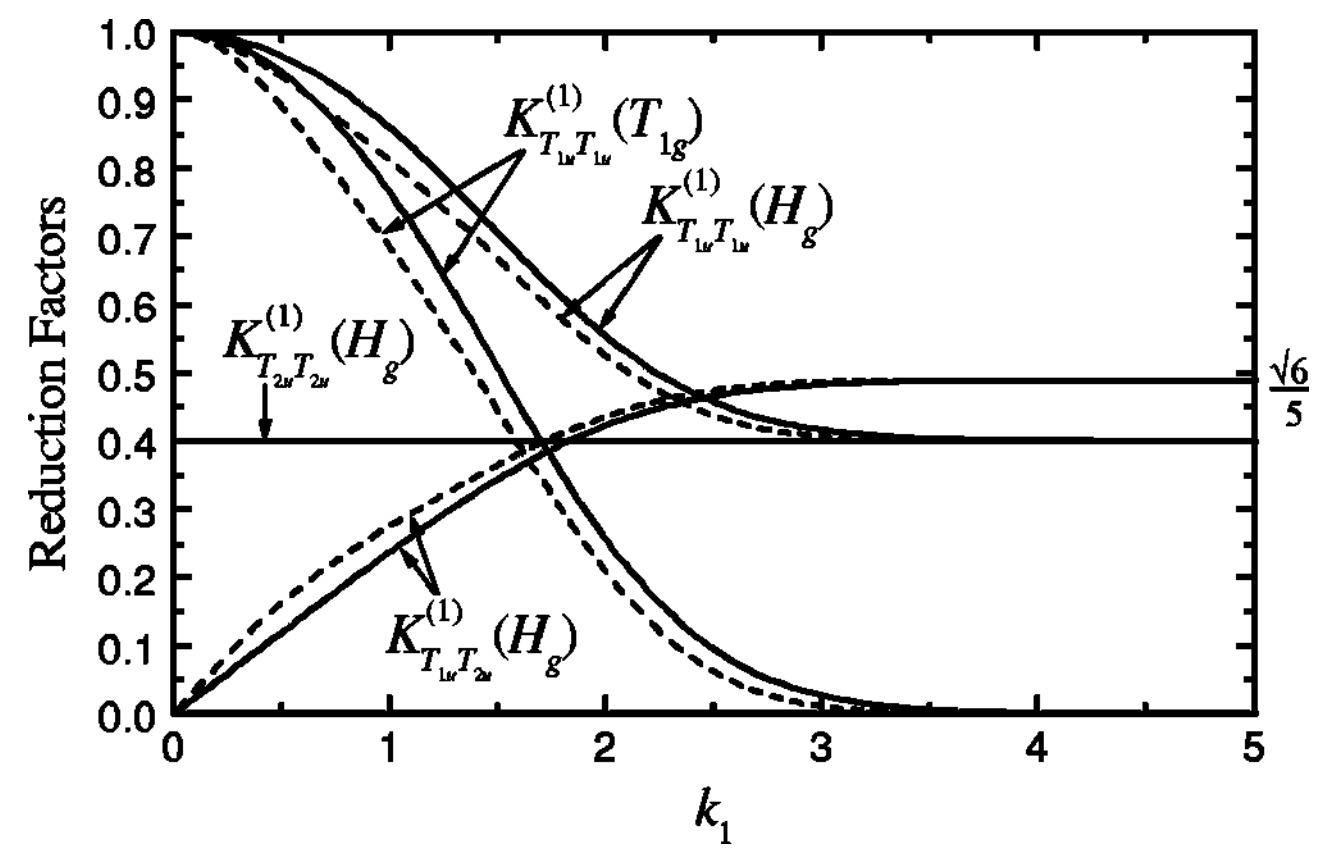

FIG. 1. A plot of the first-order RF's between symmetry-adapted states for $D_{5d}$ wells both with anisotropy (dashed lines) and without anisotropy (solid lines) as a function of $k_1$.

The off-diagonal RF's reflect, in some senses, the quenching of the strength of the interactions between two states of different symmetries. This is demonstrated in $K^{(1)}_{T_{1u}T_{2u}}(H_g)$ and $K^{(1)}_{T_{1u}G_u}(H_g)$ for trigonal wells. In weak coupling, these factors both tend to zero as the energy gap between $T_{1u}$ and $T_{2u}$ (or $G_u$) is large. On the other hand, as the RF's for $T_{1g}$ electronic operators approach zero in the strong coupling limit, $H_g$ operators dominate the energies of the JT system in strong coupling. We note that, for the $D_{5d}$ case, the RF's are similar to those for the cubic $T\otimes(e\oplus t_2)$ system with equal coupling.[61]

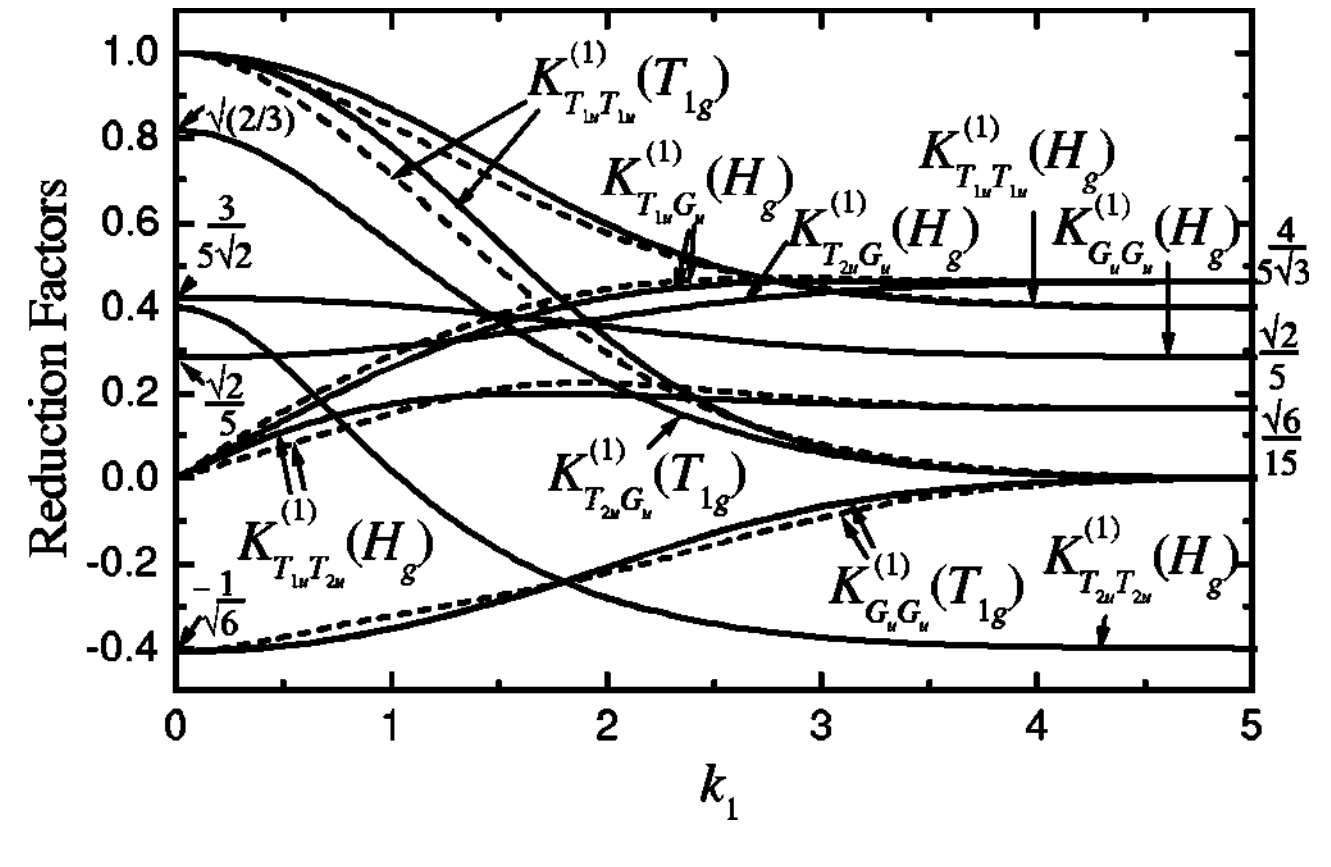

FIG. 2. As in Fig. 1 but for $D_{3d}$ wells.

Anisotropy in the wells alters the frequencies of the oscillators located in the wells.[48] Expressions for the frequencies can be determined using a scale transformation procedure in addition to the shift transformation. It is found to be also necessary to include second-order corrections to the well states in order to obtain accurate results.[52] When anisotropy is included, the simple expressions for the first-order RF's given above should be modified. However, the results are then very complex and cannot be expressed in simple analytical forms. In most cases, it is possible, as a first approximation, to neglect the second-order corrections to the well states and modify only the expressions for the overlap integrals to include anisotropy.[48] Results for the RF's where this approximation does not produce divergent results are plotted as dashed lines in Figs. 1 and 2 alongside the corresponding isotropic results (solid lines). We have taken $V_2' = -V_3' = 0.04k_1$ for $D_{5d}$ wells and $V_2' = -V_3' = -0.04k_1$ for $D_{3d}$ wells. These values both ensure that the correct wells are absolute minima and meet the condition that when $k_1$ approaches zero, $V_2'$ and $V_3'$ also approach zero. They also give $V_2'$ an approximate value to explain observed experimental data on $C_{60}$.[62] It can be seen that, as would be expected, the anisotropic corrections become negligible in strong coupling. It can also be seen that the corrections in the intermediate coupling region are relatively small. Therefore it can be concluded that the much simpler results obtained neglecting anisotropy are a reasonable approximation to the true results.

### B. Second-order reduction factors

Second-order RF's can be calculated using the general methods given previously.[37] The excited states used in the calculation should be taken to be symmetry-adapted states derived using projection operator methods. However, as discussed in Sec. I, the calculation of such states is very complicated in icosahedral symmetry, so the excited states will be taken to be the zero-phonon tunneling states and well states with phonon excitations, even though the latter are only strictly appropriate in the infinite-coupling limit. In addition, the tunneling levels require special consideration as they become degenerate with the ground states in strong coupling. Hence these will be considered separately later. Anisotropic effects will also be ignored throughout these calculations. We will only consider RF's within the ground ($T_{1u}$) state, and not ''off-diagonal'' second-order RF's involving the tunneling levels, as these are the most useful for subsequent calculations involving effective Hamiltonians.

Thus, for a purely electronic perturbation of symmetry $\Gamma$, the second-order perturbation Hamiltonian neglecting the tunneling states is

$$\mathcal{H}^{(2)}(\Gamma\otimes\Gamma) = \mathcal{H}^{(1)}(\Gamma)G(T_1)\mathcal{H}^{(1)}(\Gamma), \tag{15}$$

where

$$G(T_1) = -\sum_{n,m} \frac{|\psi^{(m)\prime};n\rangle\langle\psi^{(m)\prime};n|}{\Delta E} \tag{16}$$

and where the sum is taken over all possible phonon excitations $n$ and well states $\psi^{(m)}$. $\Delta E$ is the difference in energy between the excited well states $|\psi^{(m)\prime};n\rangle$ and the ground vibronic symmetry-adapted states.

Substituting Eq. (7) for the first-order perturbation into Eq. (15), we have

$$\mathcal{H}^{(2)}(\Gamma\otimes\Gamma)=\sum_{\gamma_j\gamma_k} W^{+}_{\Gamma\gamma_j}W_{\Gamma\gamma_k}C^{\dagger}_{\Gamma\gamma_j}G(T_1)C_{\Gamma\gamma_k}. \tag{17}$$

(Note that the dagger was omitted in our earlier paper as the matrices are real.[37] It is included here to allow replacement of the operators $C$ and $W$ by complex orbital and spin operators in Sec. III D.) From a symmetry point of view, $G(T_1)$ is a scalar and thus it does not change the transformation properties of any term in the sum. The terms within $\mathcal{H}^{(2)}$ have the same symmetry properties as those of the operator $C^{\dagger}_{\Gamma\gamma}G(T_1)C_{\Gamma\gamma}$ and they are thus second-rank tensors. They can therefore be expressed as a sum of irreducible tensors in the form

$$C^{\dagger}_{\Gamma\gamma_j}G(T_1)C_{\Gamma\gamma_k}=\sum_{M\mu}\mathcal{L}^{(2)}_{M\mu}(\Gamma\otimes\Gamma)\langle\Gamma\gamma_j\Gamma\gamma_k|M\mu\rangle, \tag{18}$$

where

$$\mathcal{L}^{(2)}_{M\mu}(\Gamma\otimes\Gamma)=\sum_{\gamma_j\gamma_k} C^{\dagger}_{\Gamma\gamma_j}G(T_1)C_{\Gamma\gamma_k}\langle\Gamma\gamma_j\Gamma\gamma_k|M\mu\rangle, \tag{19}$$

and $M$ is taken over all the elements contained in $\Gamma\otimes\Gamma$ and $\mu$ over all the components of the irreducible representation $M$.

Now, by definition, the effect of a second-order effective Hamiltonian $\mathcal{H}^{(2)}_{eff}$ between electronic states $|\Gamma_l\gamma_k\rangle$ must be the same as the effect of $\mathcal{H}^{(2)}(\Gamma\otimes\Gamma)$ between the vibronic ground states $|0\Gamma_l\gamma k\rangle$, i.e.,

$$\langle\Gamma_l\gamma_j|\mathcal{H}^{(2)}_{eff}(\Gamma\otimes\Gamma)|\Gamma_l\gamma_k\rangle\equiv\langle 0\Gamma_l\gamma_j|\mathcal{H}^{(2)}(\Gamma\otimes\Gamma)|0\Gamma_l\gamma_k\rangle. \tag{20}$$

The second-order RF's $K^{(2)}_M$ are defined to be factors multiplying electronic operators

$$L^{(2)}_{M\mu}(\Gamma\otimes\Gamma)=\sum_{\gamma_j\gamma_k} C^{\dagger}_{\Gamma\gamma_j}C_{\Gamma\gamma_k}\langle\Gamma\gamma_j\Gamma\gamma_k|M\mu\rangle \tag{21}$$

such that

$$\mathcal{H}^{(2)}_{eff}(\Gamma\otimes\Gamma)=\sum_{M\mu}\sum_{\gamma_j\gamma_k} W^{+}_{\Gamma\gamma_j}W_{\Gamma\gamma_k}\times\langle\Gamma\gamma_j\Gamma\gamma_k|M\mu\rangle K^{(2)}_M L^{(2)}_{M\mu}(\Gamma\otimes\Gamma). \tag{22}$$

Therefore, the second-order RF's, which essentially incorporate all the vibronic effects, can be written as

$$K^{(2)}_M(\Gamma)=\frac{\langle 0\Gamma_l\gamma_j|\mathcal{L}^{(2)}_{M\mu}(\Gamma\otimes\Gamma)|0\Gamma_l\gamma_k\rangle}{\langle\Gamma_l\gamma_j|L^{(2)}_{M\mu}(\Gamma\otimes\Gamma)|\Gamma_l\gamma_k\rangle}. \tag{23}$$

The second-order RF's are independent of the symmetry component labels $\mu$, $\gamma_j$, and $\gamma_k$.

The only perturbations $\Gamma$ giving nonzero second-order RF's for the $T_{1u}\otimes h_g$ system are those of $T_{1g}$ or $H_g$ symmetry. The only values of $M$ allowed are those contained in the product $\Gamma\otimes\Gamma$, namely, $A_g$, $T_{1g}$, and $H_g$. For perturbations of $H_g$ symmetry, there are two possible $H_g$-type RF's due to the repeated root in $H_g\otimes H_g$. These will be labeled $H_{1g}$ and $H_{2g}$.

The second-order RF's as given in Eq. (23) can be evaluated for any given system using the appropriate states and operators. After much algebra, it is found that (because the excited states have been approximated to simple harmonic oscillator functions localized in the potential wells) the final results are a linear combination of $m$-dimensional sums having functional forms equivalent to the two-dimensional function

$$\sum_{l,m=0}^{\infty}{}' \frac{X^l Y^m}{(E+l+m)l!m!}, \tag{24}$$

where the prime on the sum indicates that the term with $l=m=0$ is excluded and $E$ is the difference in energy between the excited phonon states and the ground states in units of $\hbar\omega$. The maximum dimension of the sum is 5, as there are five components of the $h_g$ mode. Although these factors can be computed directly, the $m$-dimensional sums can be simplified[13] to one-dimensional sums of the form

$$f(Z)=\sum_{n=1}^{\infty}\frac{Z^n}{(E+n)n!} \tag{25}$$

that are quicker to compute.

The energies of the symmetry-adapted ground states have already been calculated[48] for all coupling strengths. As the excited states are states localized in the wells, the energy of a state with $n$ phonon excitations can be obtained by taking the strong-coupling limit of the ground-state energy ($S_I$ or $S_D\to 0$) and adding $n\hbar\omega$. Therefore, neglecting quadratic coupling, the energy $E$ in the function $f(Z)$ in Eq. (25) is

$$E=\frac{X_{5d}S_I}{(1+S_I)} \tag{26}$$

for $D_{5d}$ wells and

$$E=\frac{X_{3d}S_D(5+4S_D)}{(3+5S_D+2S_D^2)} \tag{27}$$

for $D_{3d}$ wells, where $X_{5d}=12k_1^2/25$ and $X_{3d}=4k_1^2/15$. An alternative approximation is to assume that the excited states are all $n\hbar\omega$ above the ground state, as used previously for cubic systems.[45] In this case, $E$ is taken to be zero.

For $D_{5d}$ wells, the RF's can be written in terms of the functions $f_1=f(X_{5d})$ and $f_2=f(2X_{5d})$. The results are

$$K^{(2)}_{A_g}(T_{1g})=25(f_1+f_2)Y_{5d},$$

$$K^{(2)}_{T_{1g}}(T_{1g})=50f_1Y_{5d},$$

$$K^{(2)}_{H_g}(T_{1g})=10(4f_1+f_2)Y_{5d}, \tag{28}$$

and

$$K^{(2)}_{A_g}(H_g)=5(3f_1+4f_2)Y_{5d},$$

$$K^{(2)}_{T_{1g}}(H_g)=30f_1Y_{5d},$$

$$K^{(2)}_{H_{1g}}(H_g)=10(2f_1+f_2)Y_{5d},$$

$$K^{(2)}_{H_{2g}}(H_g)=2(18f_1+f_2)Y_{5d}, \qquad (29)$$

for a $T_{1g}$ and a $H_g$ perturbation, respectively, where $Y_{5d}=-2S_I^2/[25(1+S_I)]$. For $D_{3d}$ wells, the RF's can be written in terms of the functions $g_1=f(X_{3d})$ and $g_2=f(2X_{3d})$, $g_3=f(3X_{3d})$ and $g_4=f(4X_{3d})$. The results are

$$K^{(2)}_{A_g}(T_{1g})=5[3g_2+2(3g_1+5g_2)S_D+(5g_2+15g_3+12g_4)S_D^2]Y_{3d},$$

$$K^{(2)}_{T_{1g}}(T_{1g})=10S_D[2(3g_1+g_2)+(13g_2+3g_3)S_D]Y_{3d},$$

$$K^{(2)}_{H_g}(T_{1g})=2[3g_2+8(3g_1+2g_2)S_D+4(11g_2+6g_3+3g_4)S_D^2]Y_{3d}, \qquad (30)$$

and

$$K^{(2)}_{A_g}(H_g)=[24g_2+2(9g_1+25g_2)S_D+(13g_2+45g_3+42g_4)S_D^2]Y_{3d},$$

$$K^{(2)}_{T_{1g}}(H_g)=2S_D[2(21g_1+5g_2)+(23g_2+9g_3)S_D]Y_{3d},$$

$$K^{(2)}_{H_{1g}}(H_g)=[14g_2+8(5g_1+4g_2)S_D+4(17g_2+5g_3+2g_4)S_D^2]Y_{3d},$$

$$K^{(2)}_{H_{2g}}(H_g)=2[3g_2+4(9g_1+4g_2)S_D+2(g_2+9g_3+6g_4)S_D^2]Y_{3d}, \qquad (31)$$

where $Y_{3d}=-2S_D^2/[15(3+5S_D+2S_D^2)]$.

It is interesting to note that the RF's have been expressed in terms of two related functions for $D_{5d}$ wells and four related functions for $D_{3d}$ wells. This is a very similar result to that found previously for the cubic $T\otimes e$ and $T\otimes t$ systems,[45] where in both cases the RF's were expressed in terms of two functions $f(X)$ and $f(2X)$.

It is a simple matter to numerically compute the above expressions for the RF's for any given coupling strength. Sufficient phonon excitations are included in the calculation to ensure that additional contributions from higher excited states are negligibly small for the coupling strengths of interest, noting that the number of significant excited states increases as the coupling strength increases. Figures 3 and 4 show the RF's (in units of $1/\hbar\omega$) as a function of the coupling strength $k_1$ up to $k_1=5$ for $T_1$ and $H$ perturbations, respectively, using the energies $E$ given in Eqs. (26) and (27). In both figures, solid lines are used to denote the results for $D_{3d}$ wells and dashed lines the results for $D_{5d}$ wells.

It can be seen that all of the RF's are negative. Hence it is important to note that although these quantities are by convention called ''reduction factors'' they indicate both the magnitude and sign of additional terms that need to be added to effective Hamiltonians when vibronic coupling is included compared to when it is not. For each perturbation $\Gamma$, the $K^{(2)}_{A_g}(\Gamma)$ RF has the largest magnitudes and the $K^{(2)}_{T_{1g}}(\Gamma)$ RF the smallest magnitude. If the energy $E$ is taken to be zero, the curves obtained have a similar shape but are larger in magnitude in intermediate coupling. The $K^{(2)}_{A_g}(T_{1g})$ RF has a maximum magnitude of $1.65/\hbar\omega$ and the $K^{(2)}_{A_g}(H_g)$ RF a maximum magnitude of $1.31/\hbar\omega$, compared to $1.31/\hbar\omega$ and $1.03/\hbar\omega$, respectively, with the energies in Eqs. (26) and (27).

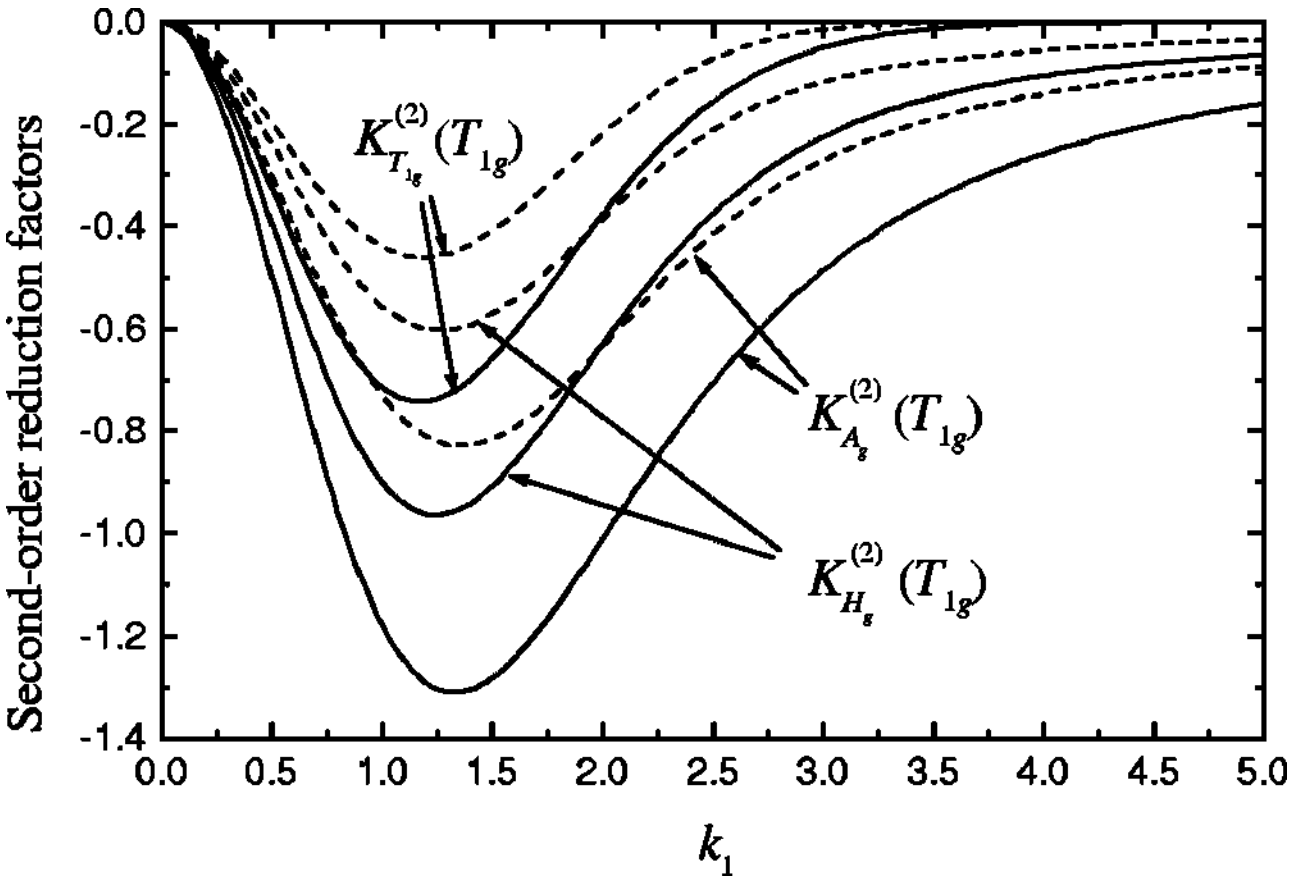

FIG. 3. Second-order RF's for a perturbation of $T_1$ symmetry. The solid lines are the results obtained using symmetry-adapted states for $D_{3d}$ wells and the dashed lines the results for $D_{5d}$ wells.

The RF's $K^{(2)}_\Gamma(T_{1g})\equiv K^{(3d)}_\Gamma(T_{1g})$ for $D_{3d}$ wells are all almost exactly 1.6 times larger than the corresponding RF's $K^{(2)}_\Gamma(T_{1g})\equiv K^{(5d)}_\Gamma(T_{1g})$ for $D_{5d}$ wells up to $k_1\simeq 2$ and also in very strong coupling. In the intermediate-coupling region, the $K^{(3d)}_\Gamma(T_{1g})$ are slightly more than $1.6K^{(5d)}_\Gamma(T_{1g})$, although the discrepancy is never more than $0.05/\hbar\omega$. For the $H$ perturbation, a similar scaling effect can be observed although the equivalence is not so exact as for the $T$ perturbation, especially in strong coupling. The $K^{(3d)}_\Gamma(H_g)$ RF's are all approximately 1.7 times larger than the corresponding $K^{(5d)}_\Gamma(H_g)$ RF's up to $k_1\simeq 2$, with the exception of the $K^{(5d)}_{H_{2g}}(H_g)$ RF. The latter has the same asymptotic behavior as the $K^{(5d)}_{T_{1g}}(H_g)$ RF in strong coupling, whereas the other three $K_{H_g}(H_g)$ RF's all have their own unique asymptotic behavior.

In the cubic $T_1\otimes t_2$ system, second-order RF's have been calculated with both symmetry-adapted and excited well states. This latter approximation was found to result in a slight overestimation of the RF's in intermediate coupling due to the inclusion of too much overlap.[59] Hence the values of the reduction factors calculated here for the intermediate-coupling region should be taken to be an upper limit on the true values.

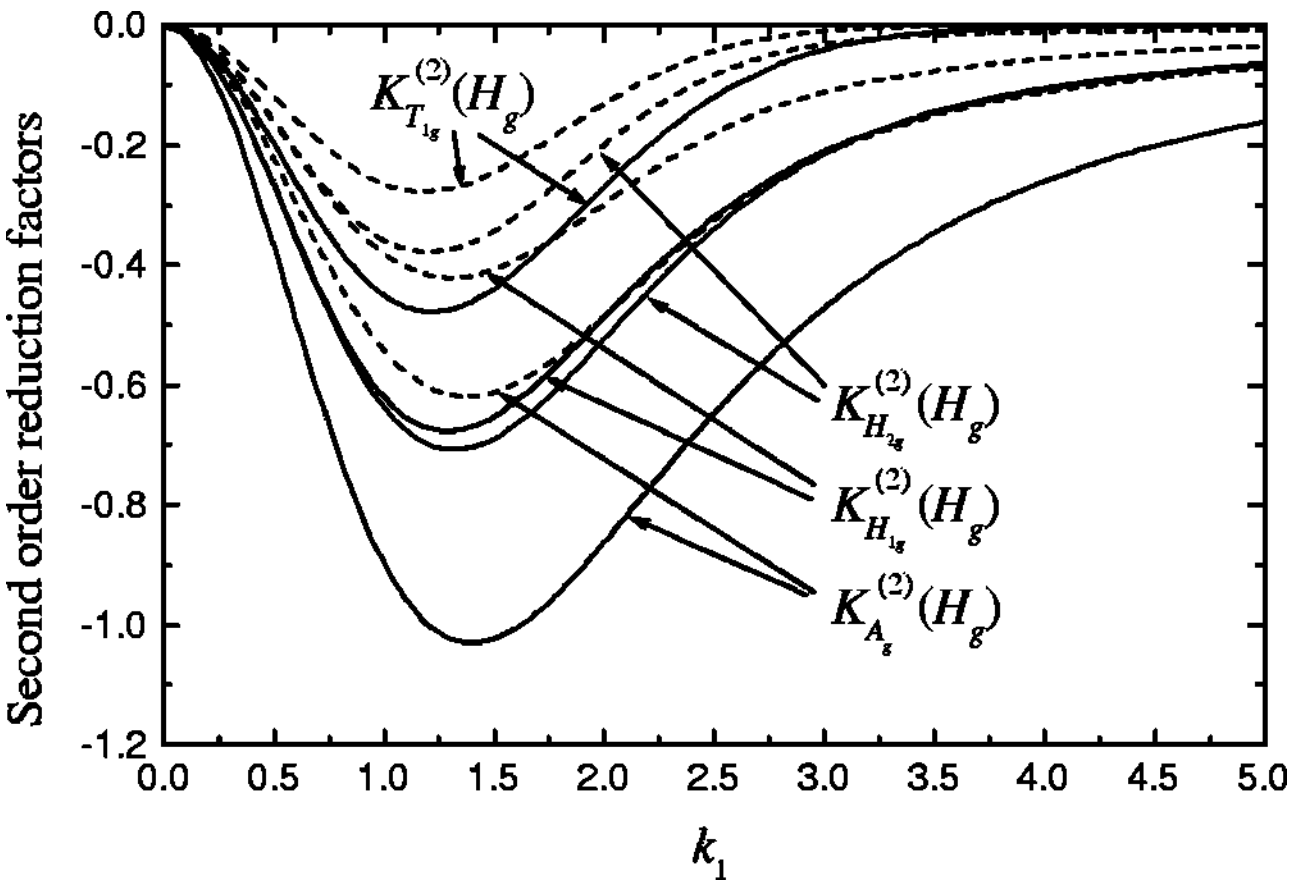

FIG. 4. As in Fig. 3 but for a perturbation of $H$ symmetry.

### C. Tunneling levels

As mentioned above, any tunneling levels that are coupled to the ground state by the perturbation in question should be considered in the calculations. In this case, no tunneling levels are coupled by the $T_{1g}$ perturbation. However, all tunneling levels are coupled for the $H_g$ perturbation. The contributions can be calculated by substituting appropriate expressions for the tunneling states instead of $|\psi^{(m)\prime};n\rangle$ into Eq. (16). If the usual symmetry-adapted states in Eqs. (5) and (6) are used, the contributions are found to rapidly diverge as the coupling strength $k_1$ increases to around 2 or 3. Indeed, tunneling contributions will always diverge when the tunneling splitting(s) and the energy differences caused by the perturbation are much smaller that $\hbar\omega$. This is because the energy difference in the denominator tends to zero and the states that are being used are not correct to first order in perturbation theory in strong coupling.

The divergence of the tunneling contributions is not a serious problem because the concept of reduction factors is not valid in the strong-coupling limit; the strongly-coupled $T_{1u}\otimes h_g$ JT system cannot be described in terms of an effective orbital triplet. For vibronic systems with tunneling levels, the correct approach in strong coupling is to work in an enlarged quasidegenerate basis include all tunneling levels. First-order RF's can then be calculated that are relevant for this basis. This results in new combinations of the symmetry-adapted states that are correct states of the system to first order in perturbation theory. In this case, second-order RF's could also be calculated that do not diverge in strong coupling.

Although the concept of RF's is not valid in strong coupling, it is still useful to estimate the sizes of the RF contributions in intermediate coupling to determine whether they are significant, or whether they can be neglected as has been done for other systems.[36] Suitable states for the strong-coupling limit are those states that simultaneously diagonalize each of the $C_{H\gamma}$ matrices in Eq. (3). These are found to be simply the zero-phonon states localized in the wells. However, although the numerator in the general expression (23) for the second-order RF's calculated using these zero-phonon well states correctly tends to zero in strong coupling, these states can not be used in place of the symmetry-adapted states because they all have the same energy and so the denominator is exactly zero at all couplings.

We know that the usual symmetry-adapted states $|0\Gamma\gamma\rangle$ in Eqs. (5) and (6) are good states for weak coupling, and that the states $|\psi^{(m)\prime};0\rangle$ localized in the wells are good states in strong coupling. We have therefore constructed combinations of the the states that we know tend to the correct strong- and weak-coupling limits, namely, states of the form

$$a|0\Gamma\gamma\rangle+b|\psi^{(m)\prime};0\rangle, \tag{32}$$

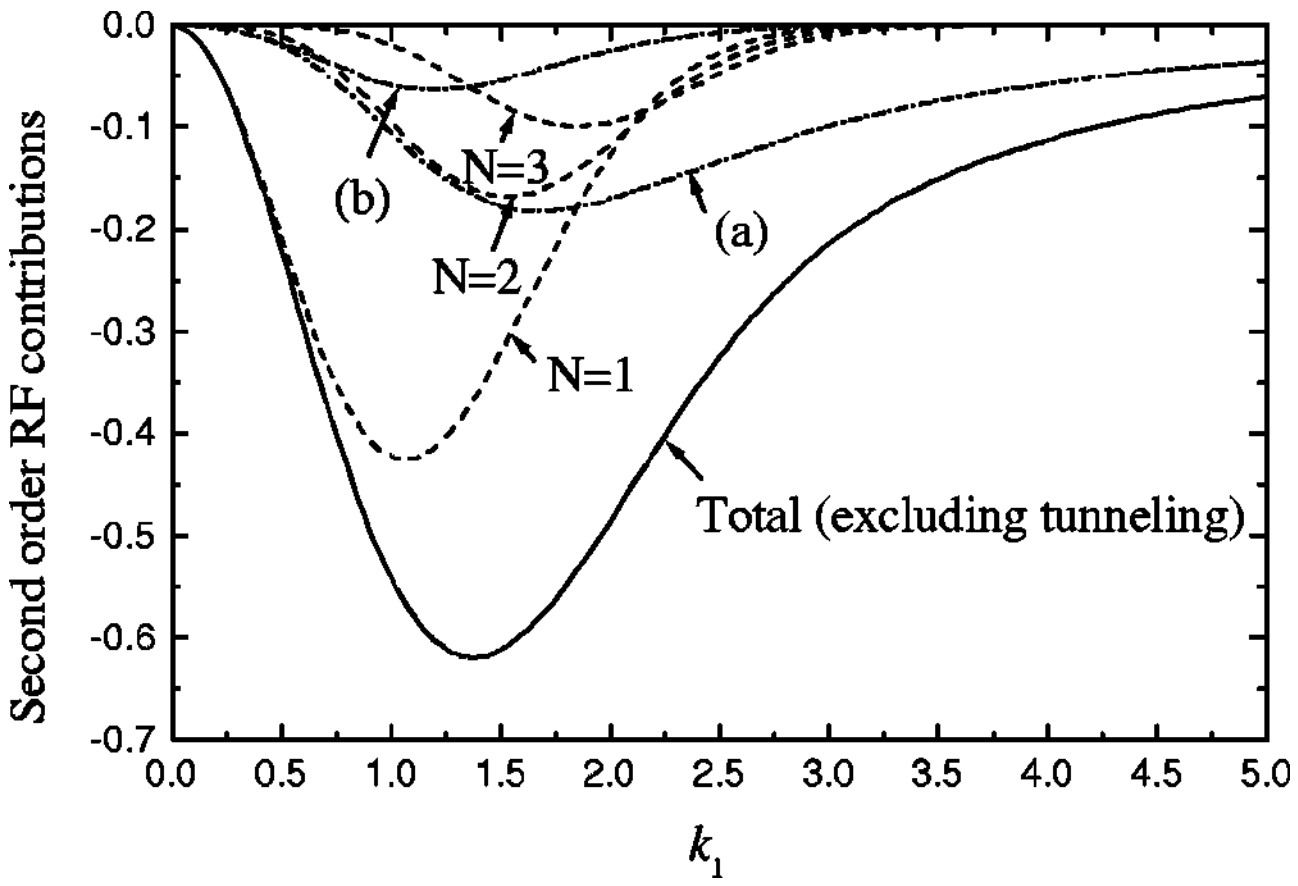

FIG. 5. The total $K^{(5d)}_{A_g}(H_g)$ RF excluding tunneling (solid line), the contributions from well states with one, two, and three phonons (dashed line) and two different estimations of the tunneling level contribution [dot-dash lines (a) and (b)].

where $a=S_I$ and $b=1-S_I$, for example. The main question is to determine which well state to associate with which symmetry-adapted state. When Eqs. (5) and (6), giving the symmetry-adapted states in terms of the well states, are solved to obtain expressions for the well states in terms of the symmetry-adapted states, either in the strong-coupling limit or at all couplings, it is found that the results naturally divide into two triplets for $D_{5d}$ wells and two triplets and a quadruplet for $D_{3d}$ wells. For example, for the $D_{5d}$ wells in strong coupling, one triplet is

$$|\psi_A\rangle=N(|0T^{(5d)}_{1uy}\rangle+\phi^{-1}|0T^{(5d)}_{1uz}\rangle-\phi^{-1}|0T^{(5d)}_{2uy}\rangle+|0T^{(5d)}_{2uz}\rangle),$$

$$|\psi_C\rangle=N(|0T^{(5d)}_{1uz}\rangle+\phi^{-1}|0T^{(5d)}_{1ux}\rangle-\phi^{-1}|0T^{(5d)}_{2uz}\rangle+|0T^{(5d)}_{2ux}\rangle),$$

$$|\psi_E\rangle=N(|0T^{(5d)}_{1ux}\rangle+\phi^{-1}|0T^{(5d)}_{1uy}\rangle-\phi^{-1}|0T^{(5d)}_{2ux}\rangle+|0T^{(5d)}_{2uy}\rangle), \tag{33}$$

corresponding to well states *A*, *C*, and *E* respectively, and the other triplet is

$$|\psi_B\rangle=N(\phi^{-1}|0T^{(5d)}_{1uz}\rangle-|0T^{(5d)}_{1uy}\rangle+|0T^{(5d)}_{2uz}\rangle+\phi^{-1}|0T^{(5d)}_{2uy}\rangle),$$

$$|\psi_D\rangle=N(\phi^{-1}|0T^{(5d)}_{1ux}\rangle-|0T^{(5d)}_{1uz}\rangle+|0T^{(5d)}_{2ux}\rangle+\phi^{-1}|0T^{(5d)}_{2uz}\rangle),$$

$$|\psi_F\rangle=N(\phi^{-1}|0T^{(5d)}_{1uy}\rangle-|0T^{(5d)}_{1ux}\rangle+|0T^{(5d)}_{2uy}\rangle+\phi^{-1}|0T^{(5d)}_{2ux}\rangle), \tag{34}$$

corresponding to wells *B*, *D*, and *F*, where $N=[2(1+\phi^{-2})]^{-1/2}$. We choose one of the triplets to correspond to the $|0T_{1u}\rangle$ states and the other with the $|0T_{2u}\rangle$ states (and the quadruplet with the $|G_u\rangle$ states for $D_{3d}$ wells). We start by associating each symmetrized state with the strong-coupling state from the nominated multiplet that contains the largest coefficient of that symmetrized states, such as $|\psi_E\rangle$ with $|0T^{(5d)}_{1ux}\rangle$.

It is found that the RF contributions calculated as described above do converge in strong coupling. Figure 5 shows the tunneling contribution [labeled (a)], together with the total $K^{(5d)}_{A_g}(H_g)$ RF, neglecting tunneling. Also shown are the contributions to the RF from the well states with one-, two- and three- phonon excitations. It is found that for all of the RF's, the calculated tunneling contribution up to $k_1 \simeq 2$ is similar in size to the contribution of the two-phonon well states, but that in stronger coupling the calculated tunneling contribution is of the same order of magnitude as the overall contribution neglecting tunneling. However, this result depends upon the method of calculation rather than a true physical result. The procedure adopted has shown that the divergence can be removed by redefining the basis states; however, it does not give any information on the *rate* of convergence in strong coupling. Indeed, such information is also not relevant because, as mentioned above, the whole RF formalism is not appropriate in strong coupling.

To illustrate the above point further, an alternative set of basis states is defined in which the strong-coupling parts are not simply single well states but a linear combination of the well states from the nominated multiplet. The coefficients are chosen so that the coefficients of the symmetrized states constructively add, such as $(|\psi_C\rangle + |\psi_E\rangle)$ or $(|\psi_D\rangle - |\psi_F\rangle)$ with $|0T_{1ux}^{(5d)}\rangle$. The result is given as line (b) in Fig. 5. This time, it can be seen that the RF is much smaller for all coupling strengths and has decayed to zero almost completely by $k_1 \simeq 2.5$. With this choice, the tunneling contributions only serve to marginally increase the value of the maximum magnitude and have no effect in strong coupling.

We have also investigated alternative situations with different associations between the strong- and weak-coupling states, and with different values of $a$ and $b$. In all cases, a guide to whether the choice of states is good or bad is to evaluate the tunneling splittings and compare them with the tunneling splittings using the symmetry-adapted states alone. If a tunneling splitting is negative for some coupling strengths, that case is rejected as a poor choice. It is found that for some choices of states the results converge, and in others they diverge. Where convergence occurs, the results obtained with different choices of states all show the same behavior as in Fig. 5. Although the actual numerical values are different in different cases, they still have the same orders of magnitude as each other. The results are not sensitive to which triplet is associated with $T_{1u}$ and which with $T_{2u}$.

The net conclusion of the calculations involving the tunneling contribution is that, as expected, the contributions do converge to zero in infinite coupling when correct strong-coupling states are taken. However, it is not possible to obtain precise numerical values for the tunneling contributions. It has been shown likely that the tunneling contributions to the RF's can be neglected in weak and moderate couplings, in line with the findings in other systems.[36] In strong-coupling, the concept of RF's in the basis of an electronic triplet is not appropriate and an alternative approach to the modeling of the vibronic system must be found.

### D. Example of application: Spin-orbit coupling

The first-order RF's given by Eqs. (13) and (14), and the second-order RF's given by Eqs. (28) to (31), as displayed in Figs. 1 to 4, give all the information necessary to express an effective Hamiltonian [Eqs. (9) and (22)] for any given perturbation in terms of the coupling strengths and frequencies only. The only other details required to write down the Hamiltonian explicitly are the CG coefficients given by Fowler and Ceulemans.[49] Thus all the information necessary to model spectroscopic date has already been obtained. However, as the formulation is somewhat mathematical, it is useful to illustrate how the results can be used by means of a simple example. For this purpose, we will consider the effect of spin-orbit coupling on the ground state of the vibronic $T_{1u} \otimes h_g$ JT system.

Spin-orbit coupling transforms with symmetry $\Gamma = T_1$. From Table 2 of Fowler and Ceulemans,[49] it can be seen that $C_{T_1\gamma} = L_\gamma / i\hbar\sqrt{2}$ ($\gamma = x,y,z$), where the $L_\gamma$ are the usual angular momentum operators for $l=1$ (i.e., $L_x = yP_z - zP_y$, etc.). The $W_{T_1\gamma}$ are spin operators having the same transformation properties as the $C_{T_1\gamma}$. If the overall constant is chosen so that the real Hamiltonian (7) takes the usual form $\lambda \mathbf{L}\cdot\mathbf{S}$, it follows that the first-order effective Hamiltonian is

$$\mathcal{H}_{eff}^{(1)} = \lambda K_{T_{1u}T_{1u}}^{(1)}(T_1)\mathbf{L}\cdot\mathbf{S}. \tag{35}$$

Choosing the same overall constant and using Eq. (22), it follows that the second-order effective Hamiltonian is

$$\mathcal{H}_{eff}^{(2)} = \lambda^2 \left( K_A^{(2)} L_A^{(2)} S_A^{(2)} + K_{T_1}^{(2)} \sum_\gamma L_{T_1\gamma}^{(2)} S_{T_1\gamma}^{(2)} + K_H^{(2)} \sum_\gamma L_{H\gamma}^{(2)} S_{H\gamma}^{(2)} \right), \tag{36}$$

where

$$L_A^{(2)} = \frac{1}{\sqrt{3}}(L_x^\dagger L_x + L_y^\dagger L_y + L_z^\dagger L_z),$$

$$L_{T_1x}^{(2)} = \frac{1}{\sqrt{2}}(L_y^\dagger L_z - L_z^\dagger L_y),$$

$$L_{H\theta}^{(2)} = \frac{\phi^{-1}}{\sqrt{2}} L_x^\dagger L_x - \frac{\phi}{\sqrt{2}} L_y^\dagger L_y + \frac{1}{2} L_z^\dagger L_z,$$

$$L_{H\epsilon}^{(2)} = \frac{\phi^2}{2\sqrt{3}} L_x^\dagger L_x - \frac{\phi^{-2}}{2\sqrt{3}} L_y^\dagger L_y - \frac{1}{2}\sqrt{\frac{5}{3}} L_z^\dagger L_z,$$

$$L_{Hx}^{(2)} = \frac{1}{\sqrt{2}}(L_y^\dagger L_z + L_z^\dagger L_y), \tag{37}$$

and where the $L_{\Gamma y}^{(2)}$ and $L_{\Gamma z}^{(2)}$ can be obtained by cyclic permutation from $L_{\Gamma x}^{(2)}$ ($\Gamma = T_1$ or $H$). The $S_{\Gamma\gamma}^{(2)}$ can be obtained from the $L_{\Gamma\gamma}^{(2)}$ by replacing the orbital operators $L_\gamma$ by equivalent spin operators.

From Fig. 1, it can be seen that when the coupling is such that $D_{5d}$ wells are lowest, the first-order RF $K_{T_{1u}T_{1u}}^{(1)}(T_1)$ varies from 1 in the weak-coupling limit to 0 in the strong-coupling limit. Thus if the vibronic coupling is strong, the effect of the coupling will be to significantly quench the effect of first-order spin-orbit coupling. This is very similar to many of the cubic systems studied previously (as stated in the Introduction). However, Fig. 2 shows that if $D_{3d}$ wells are lowest, the minimum value of the first-order RF is $\sqrt{6}/15$. Thus first-order spin-orbit coupling can only ever be partially quenched, even in a very strongly coupled system. Note that if fitting to experimental data indicates that the first-order RF is less than $\sqrt{6}/15$, we can deduce that the system favors $D_{5d}$ minima. The value of the RF could then be used to estimate a value for the linear coupling strength $V_1$ and sets limits on the possible values for the quadratic couplings $V_2$ and $V_3$. If the first-order RF is predicted to be larger than $\sqrt{6}/15$, then both the $D_{3d}$ and $D_{5d}$ situations must be considered as possibilities. Any fitted values for the second-order RF's ($K_A^{(2)}$, $K_{T_1}^{(2)}$ and $K_H^{(2)}$) would further aid a determination of values for the coupling constants.

## IV. CONCLUSIONS

The main purpose of this paper has been to derive analytical expressions for the first- and second-order RF's for the $T_{1u}\otimes h_g$ JT system. These factors have been calculated using symmetry-adapted ground states and excited states located in the $D_{5d}$ and $D_{3d}$ minima. In addition, off-diagonal first-order RF's have also been derived, and corrections to the first-order RF's due to anisotropy in the potential wells obtained. The results obtained can be used to determine the parameters appearing in effective Hamiltonians used to model icosahedral systems with vibronic coupling.

As stated earlier, the ground state of a $C_{60}^-$ molecule is an electronic $T_{1u}$ triplet coupled to eight $h_g$ vibrational modes. Although only one mode is considered here, the multiple mode problem can be formulated in terms of a dominant interacting mode.[63] The remaining modes are coupled relatively weakly, and their effect can be included as a perturbation of required. Hence, it may be possible to apply the results presented here directly to the real $C_{60}^-$ problem. For example, if the results of spectroscopic experiments on $C_{60}^-$ could be modeled in terms of effective Hamiltonians, the experimental values obtained for the coefficients could be equated to the first- and second-order RF expressions obtained in this paper and hence an estimate of the strength of the vibronic coupling obtained. Currently, the evidence is that a proper interpretation of the available experimental data must involve vibronic coupling in order to explain the spectral structure and widths.[64,65] However, the strength of the coupling is still a matter of some debate. Long progressions indicating emission into high-energy vibrational levels deep in potential wells suggest the coupling is intermediate or strong.[66] However, the uv spectrum of $C_{60}^-$ in solution is similar to $C_{60}$, which could indicate that the coupling is weak.[67] The situation is complicated because it is likely that some low-symmetry perturbations are present that can stabilize static distortions away from $I_h$ symmetry.[68] RF's may also be of use in modeling these situations.

The theory for RF's developed here can be applied to other icosahedral systems, such as $G\otimes g$, $G\otimes h$, and $H\otimes g$. This has relevance to other states of $C_{60}$ and related fullerenes, such as the cation $C_{60}^+$.

## ACKNOWLEDGMENTS

The authors wish to thank the late Dr. M. C. M. O'Brien, Dr. Y.-M. Liu, and Dr. C. P. Moate for many helpful discussions. One of us (Q.C.Q.) would like to thank the U.K. committee of Vice-Chancellors and Principals and the University of Nottingham for financial support.

---

*Present address: Department of Chemistry, University of Leuven, Celestijnenlaan 200F, B-3001 Leuven, Belgium.